\documentclass[12pt,a4paper]{article}
\usepackage{graphics}
\usepackage{cite,a4wide}
\newcommand{\beq}{\begin{equation}}
\newcommand{\eeq}{\end{equation}}
\newcommand{\bea}{\begin{eqnarray}}
\newcommand{\eea}{\end{eqnarray}}
\renewcommand{\d}{\delta}
\renewcommand{\l}{\lambda}

\renewcommand{\b}{\beta}

\renewcommand{\ni}{\noindent}
\newcommand{\n}{\nu}
\newcommand{\m}{\mu}

\newcommand{\s}{\sigma}
\newcommand{\w}{\omega}

\newcommand{\A}{{\cal A}}
\renewcommand{\P}{{\cal P}}
\newcommand{\tP}{\tilde{P}}
\newcommand{\z}{{\cal Z}}

\renewcommand{\th}{\theta}

\newcommand{\vph}{\varphi}
\newcommand{\oh}{\frac{1}{2}}
\newcommand{\oq}{\frac{1}{4}}
\newcommand{\ot}{\frac{3}{2}}
\newcommand{\of}{\frac{5}{2}}
\newcommand{\dg}{\dagger}
\newcommand{\non}{\nonumber}

\newcommand{\rf}[1]{(\ref{#1})}
\newcommand{\ra}{\rightarrow}
\newcommand{\pa}{\partial}

\newcommand{\ran}{\rangle}
\newcommand{\lan}{\langle}
\newcommand{\cD}{{\cal D}}

\begin{document}

\hfill April 1998

\begin{center}

\vspace{32pt}

  {\Large \bf Center Disorder in the 3D Georgi-Glashow Model }

\vspace{36pt}

{\sl J. Ambj{\o}rn and J. Greensite}

\vspace{36pt}

  The Niels Bohr Institute \\
  Blegdamsvej 17 \\
  DK-2100 Copenhagen \O, Denmark

\vspace{120pt}

{\bf Abstract}

\end{center}

\bigskip

   We present a number of arguments relating magnetic disorder to
center disorder, in pure Yang-Mills theory in D=3 and D=4 
dimensions.  In the case of the D=3 Georgi-Glashow model, we point 
out that the abelian field distribution is not adequatedly represented, 
at very large scales, by that of a monopole Coulomb gas.  The onset of 
center disorder is associated with the breakdown of the Coulomb gas
approximation; this scale is pushed off to infinity in the $QED_3$ limit 
of the 3D Georgi-Glashow model, but should approach the color-screening 
length in the pure Yang-Mills limit.

\bigskip

\vfill

\newpage

\section{Introduction}

   The $Z_N$ center of an $SU(N)$ gauge group is associated with
the confinement properties of of a pure gauge theory in a number of ways.  
It is well known that the finite temperature confinement/deconfinement 
transition can be regarded as the breaking of a global $Z_N$ symmetry
in a volume with a compactified time direction.  In addition,
as shown by 't Hooft \cite{tHooft}, the VEV of a $Z_N$ vortex creation
operator can be interpreted as an order parameter for confinement, dual 
to Wilson loops, in $SU(N)$ gauge theory.  It was also suggested many years 
ago that ``thick'' $Z_N$ vortices are responsible for the area-law falloff 
of Wilson loops \cite{tHooft,Mack,Cop,jan,jo1,jo2,Vinci,Corn1,Feyn}, and 
recently there have been a number of numerical investigations which support 
this idea \cite{Obs,TK,LR}.

   The notion that confining (``magnetic'') disorder is center disorder may
also be supported by some simple observations, presented in section 2, 
regarding the behavior of holonomy probability distributions in Yang-Mills 
theory.  We point out that the holonomy distribution approaches a random 
distribution on the group manifold as loop size increases; however, the 
approach to the random distribution is far more rapid among the center 
elements than among elements 
of the coset.  We also show, with the help of the lattice strong-coupling 
expansion, that while center elements within a large area fluctuate 
independently,  this is not true of fluctuations in the coset for $D>2$ 
dimensions.

   The 3D Georgi-Glashow model ($GG_3$) is interesting in this context for
several reasons.  The confinement mechanism in this theory is believed
to be essentially that of compact $QED_3$, at least in some region of
the coupling parameters; one therefore expects that confining disorder 
is $U(1)$ disorder.  There is only one phase in $GG_3$, but there are two
special limits:  Compact $QED_3$ is obtained
in the limit where the mass of the W-boson becomes infinite, while 
pure Yang-Mills theory ($YM_3$) is obtained in the limit where the 
adjoint scalar effectively decouples from the gauge field. 
Since the 3D Georgi-Glashow model interpolates smoothly between $QED_3$ and 
$YM_3$, a natural question to ask is what happens to center 
disorder in $GG_3$ as we move away in parameter space from the pure 
Yang-Mills limit. 
  
   This question is taken up in section 3, where we point out a 
qualitative difference between confinement in compact $QED_3$, and 
confinement in the 3D Georgi-Glashow model at large scales.
In the case of compact $QED_3$, we show via saddlepoint methods 
that double-charged loops have twice the
string-tension of single-charged loops, while for $GG_3$, the double
(abelian) charged loops must ultimately be screened by massive W-bosons.
As a consequence, the effective abelian theory corresponding to the 3D
Georgi-Glashow model, obtained after integrating out the charged bosons,
is not adequately represented by a Coulomb gas of 't Hooft-Polyakov 
monopoles.  Our main point is that the massive W-bosons of $GG_3$ are not just 
spectators whose effect on vacuum fluctuations, beyond the range $M_W^{-1}$,
is negligible; in fact the W-bosons must strongly affect the vacuum 
distribution of abelian flux at  large distance scales.  At these
large scales, it appears that confining disorder in $GG_3$, as in the pure 
Yang-Mills theory, is associated with $Z_2$ (rather than $U(1)$) disorder.
In the $QED_3$ limit, the onset of $Z_2$ disorder is pushed off to 
infinity, while in the pure Yang-Mills limit, it roughly coincides with
the onset color screening.

\section{Confining Disorder as Center Disorder}

   A Wilson loop is understood as measuring the response of the vacuum
to the introduction of heavy sources, but it can also be viewed as
providing information about field fluctuations in the ground state, in
the absence of external charges.  Consider, in particular, a gauge theory
with matter fields in the fundamental representation.  The asymptotic
perimeter-law falloff of the Wilson loop is explained by the binding
of matter quanta to the external charge, forming a color singlet.  On
the other hand, imagine integrating out the matter fields, leaving an
effective action involving only the gauge fields.  It is then clear that
the effect of the virtual matter fields is to modify the probability
distribution of gauge-field fluctuations, such that confining configurations,
which would normally induce an asymptotic area-law falloff in the loop,
are suppressed.

   In discussing the probability distribution of gauge fields, with or
without the presence of matter fields,  it will be helpful to introduce a
gauge-invariant operator which is somewhat more general than a Wilson loop.
Consider, for simplicity, a lattice pure-gauge theory with an SU(2) gauge 
group, and let $U(C)$ denote the path-ordered product of link variables along
a closed loop $C$ (the holonomy).  The expectation value
\bea
        P_C[g]  &=& <\d[g,U]> = <\sum_j \chi_j[g] \chi_j[U(C)]>
\non \\
                &=& 1 + \sum_{j\ne 0} W_j(C) \chi_j[g]
\label{prob}
\eea
is the probability density, on the SU(2) group manifold, that the loop 
$U(C)$ equals the group element $g$.  The sum over $j$ runs over group
representations, the $\chi_j(g)$ are SU(2) group characters, and 
$W_j(C)=<\chi_j[U(C)]>$ is the VEV of the Wilson loop in representation
$j$.  As loop $C$ becomes large, this holonomy probability distribution
approaches the random distribution, $P_C(g) \ra 1$, and it will be
useful to focus on the deviation of $P_C$,  denoted $\tP_C[g]$, from the 
random distribution
\beq
        \tP_C[g] = P_C[g] - 1 
\eeq
Since $P_C[g]$ is gauge-invariant, it can only depend
on the eigenvalues of the unitary matrix $g$, and $P_C$ has flat directions
on the group manifold corresponding to $g \ra g'= u g u^\dg$.  In the 
particular case of $SU(2)$, $P_C$ only depends on $\mbox{Tr}(g)$, and we will
be interested in how this dependence fades away as the loop $C$ becomes 
large.

   A Wilson loop $W_j(C)$ can be thought of as a moment of the probability
distribution $P_C(g)$.  It is expected that, in $D=3$ and $D=4$ dimensions,
planar Wilson loops have the asymptotic form
\beq
       W_j(C) = \left\{ \begin{array}{cl}
          \exp[- \s \A(C) - \m_j \P(C) - c_j] & j = \mbox{half-integer} \cr
          \exp[- \m_j \P(C) - c_j] & j = \mbox{integer} 
          \end{array} \right.
\eeq
at any lattice coupling.
Note that all half-integer representations have the same string tension,
and all integer representations have zero string tension.  This is
a well-known consequence of color-screening, which seems (from numerical
studies) to set in somewhat after the onset of confining behavior.
The perimeter term reflects both short-range, perturbative contributions, 
roughly proportional to the quadratic Casimir $j(j+1)$, and also, for
$j>\oh$, the bound-state energy of gluons required to screen the color charge
to its minimum value (either $j=0$ or $j=\oh$).  The constant
$c_j$, which increases with $j$, can be attributed to the rapid initial falloff
of higher-representation loops in the so-called ``Casimir-scaling'' region,
before the onset of color-screening \cite{Cas,Cas1}.  At the point where
the asymptotic behavior sets in, the higher-representation loops have already
fallen to a rather small value as compared to lower-representation loops,
and this fact is accounted for in the constant $c_j$.  Since both 
$\m_j$ and $c_j$ increase with $j$, it follows that for large loops
\beq
      W_{1/2}(C) \gg W_{3/2}(C) \gg W_{5/2}(C) \gg ...
\label{Whalf}
\eeq
and
\beq
      W_1(C) \gg W_2(C) \gg W_3(C) \gg ...
\label{Wint}
\eeq
It also follows, for sufficiently large loops, that
\beq
      W_1(C) \gg W_{1/2}(C)
\label{W12}
\eeq
since the rhs falls off asymptotically with area-law behavior, and the
lhs only falls off with the perimeter law.\footnote{It should be
noted that condition \rf{W12} has yet to be verified numerically, at least
at zero temperature.  The onset of color screening appears to be at the 
edge, or perhaps just beyond, the range of current numerical 
simulations (cf.\ Michael in ref.\ \cite{Cas}).  The condition \emph{can} 
be verified at small $\beta$, using the lattice strong-coupling expansion.}

   Using [\ref{Whalf}-\ref{W12}], we have the leading behavior
\beq
      \tP_C(g) \approx \chi_1(g) \exp[-\m_1 \P(C) - c_1] 
\eeq
and the approach to the purely random distribution follows a perimeter-law,
rather than the area law which might have been expected.  In contrast,
in D=2 dimensions
\bea
       W^{2D}_j(C) &=& (2j+1) \exp[-\s_j \A(C)]
\non \\
         \s_j &=& -\log{I_{2j+1}(\b) \over I_1(\b)}
\eea
and therefore, asymptotically,
\beq
      \tP^{2D}_C(g) \approx  \chi_{1/2}(g) \exp[-\s_\oh \A(C)]
\eeq
This 2D distribution, unlike the 3D and 4D distributions, approachs the 
random value via an area-law falloff. 

   There is, however, a hidden area-law approach to randomness also
in the $D=3,4$ group distributions.  Let us extract a center element
from the holonomies
\beq
       \z[U(C)] =  \mbox{signTr}[U(C)]
\eeq
and ask for the probability $P_C(z)$ that $z=\z[U(C)]$, where $z=\pm 1$.
This is given by
\beq
        P_C(z) = <\oh(1 + z \times \mbox{signTr}[U(C)])>       
\eeq
The pure-random value is $P_C(z) = 1/2$, and again
we remove this constant to define the deviation 
$\tP_C(\pm) = P_C(\pm) - \oh$ from pure-random.
Then, from the character expansion
\bea
  \mbox{signTr}[U(C)] &=& \sum_{j=\oh,\ot,\of...} a_j \chi_j[U(C)]
\non \\
           a_j &=& \int dg ~  \mbox{signTr}[g] \chi_j(g)
\label{ce}
\eea
and again applying \rf{Whalf}, we find
\beq
  \tP_C(z) \approx z {8 \over 3\pi} \exp[-\s_{1/2} \A(C) -
                                          \m_{1/2}\P(C) -c_{1/2}]
\eeq
    
   The conclusion is that, although the overall holonomy 
probability distribution
$P_C[g]$ approachs the random value via a perimeter falloff in D=3 and 
D=4 dimensions, the probability that $\mbox{Tr}[U(C]$ has one or the other 
sign approachs the random distribution via an area-law falloff.  
In D=2 dimensions there is no such distinction between $P_C(g)$
and $P_C(z)$; both probabilities have an area-law falloff.  The strong 
implication is that fluctuations in the center element, which distinguishes 
between two cosets of the group characterized by the sign of Tr$(g)$, 
are the fluctuations characteristic of confining disorder in D=3 and D=4 
dimensions.  To go further, however, we will need to 
resort to the lattice strong-coupling expansion.  

   We note in passing that one finds the
relation
\beq
       <\mbox{signTr}[U(C)]> \approx  {8 \over 3\pi} W_{1/2}(C)
\eeq
from the character expansion \rf{ce}, as a consequence of the inequalities
\rf{Whalf}.  This explains the rather mysterious equality of potentials 
extracted (i) from the Wilson loops $W_{1/2}(C)$;  and (ii) from the sign 
of Wilson loops $<\mbox{signTr}[U(C)]>$, which was found recently in numerical 
simulations \cite{TK}.

\subsection{$\mathbf{Z_2}$ Disorder at Strong Coupling}       

   It is often said that confinement in strong-coupling lattice gauge
theory is simply a matter of plaquette disorder:  Group elements associated
with loops around nearby areas (the plaquettes) fluctuate independently,  
leading to an area law falloff for the Wilson loops.  This is certainly
a correct statement of the situation in $D=2$ dimensions.  However, as
we will now show, there are some important qualifications to be made 
in higher dimensions.     

\begin{figure}[t]
\centerline{\scalebox{.4}{\rotatebox{270}{\includegraphics{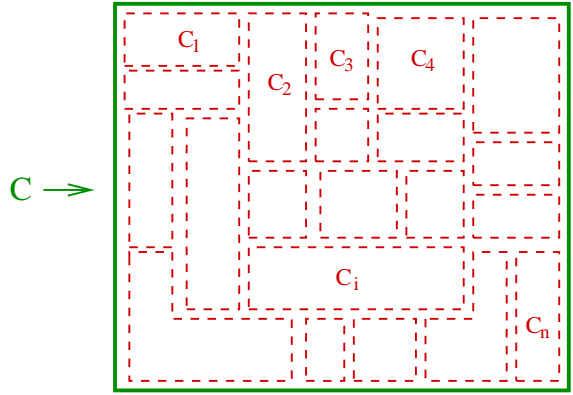}}}}
\caption{Subdivison of a large loop $C$ into smaller regions bounded by
loops $\{C_i\}$.  Note that the sum of interior perimeters may be much
greater than the total perimeter of $C$.}
\label{loops}
\end{figure}

   Let us consider a very large planar loop $C$ whose minimal area 
$\A(C)$ is subdivided into some number $n$
of subareas $\A(C_i)$, encircled by loops $C_i$ (Fig. \ref{loops}).  
If all loops $C_i$
are large, and the coupling is strong, the question is to what extent
the holonomies $U(C_i)$ fluctuate independently.  Let $F[g]$ denote
any class function with character expansion
\beq
      F[g] = \sum_{j\ne 0} f_j \chi_j(g)
\eeq
The test for whether the $U(C_i)$ fluctuate independently is
whether or not the VEV of the product of $F[U(C_i)]$ equals the
product of the VEVs, i.e.
\beq
      <\prod_i F[U(C_i)]> \stackrel{?}{=} \prod_i <F[U(C_i)]> 
\eeq
In D=2 dimensions, it is easy to verify that the equality holds
\bea
       <\prod_{i=1}^n F[U(C_i)]> &\approx& \left(\prod_i 2 f_{1/2} 
                 \right) \exp\left[-\s_{1/2} \A(C)\right]
\non \\
   &=& \prod_{i=1}^n <F[U(C_i)]>
\eea
so the group elements $U(C_i)$ do seem to fluctuate independently in
each subregion.

   In $D>2$ dimensions the answer is different.  Suppose that
each $C_i$ satisfies \\
$\A(C_i)\gg 2\P(C_i)$ in lattice units.  In that case, 
we find from the strong-coupling expansion the leading contribution to the 
VEV of products\footnote{We neglect here certain sub-leading, shape-dependent
terms in the exponent}
\bea
    <\prod_{i=1}^n F[U(C_i)]> &\approx& 
         \left({1\over 3}\right)^{n-1} f_1^n <\chi_1[U(C)]>
\non \\
     &=&   \left({1\over 3}\right)^{n-1} f_1^n \exp[-\m \P(C)]
\label{F1}
\eea
where $\m = 4 \s_{1/2}$.  The powers of $1/3$ are due to the fact that
this contribution is highly non-planar, and would vanish in the large-N
limit.  The planar contribution, however, has an area law falloff, and for
any finite N it is negligible for large loops.  Likewise, for the product
of VEVs, we have the leading contribution
\bea
    \prod_{i=1}^n <F[U(C_i)]> &\approx& 
        \prod_i f_1 <\chi_1[U(C_i)]>
\non \\
     &=&   f_1^n\exp\left[-\m \sum_i \P(C_i)\right]
\label{F2}
\eea
However, for $n\gg 1$, 
\beq
      \sum_{i=1}^n \P(C_i) \gg \P(C)
\label{F3}
\eeq
from which we see that the exponential falloff of  
$<\prod_i F[U(C_i)]>$ and $\prod_i <F[U(C_i)]>$ is quite
different, already at the leading terms in the exponents.
The conclusion is that the group elements $U(C_i)$ do not, in fact,
fluctuate independently; the deviations from pure random in the set of
sub-loop probability distributions must be correlated.

   On the other hand, since the area-law falloff of Wilson loops is supposed
to be due to magnetic disorder; \emph{some} component of magnetic
flux should be fluctuating (nearly) independently.  It is easy to
check, at strong coupling, that the center elements $\z[U(C_i)]$
have the required property.  The leading behavior for the VEV
of the product is given by 
\bea
     <\prod_i \z[U(C_i)>  &=&  <\prod_i \mbox{signTr}[U(C_i)]>
\non \\
     &\approx& <\prod_i \left(3\over8\pi\right) \chi_{1/2}[U(C_n)]>
\non \\
     &\approx& \exp[-\s_{1/2} \A(C)] \left(3\over 4\pi\right)^n 
\label{A1}
\eea
Similarly, for the product of the VEVs
\bea
     \prod_i <\z[U(C_i)>  &=&  \prod_i <\mbox{signTr}[U(C_i)]>
\non \\
 &\approx& \prod_{i=1}^n \left(3\over 4\pi\right) \exp[-\s_{1/2} \A(C_i)]
\label{A2}
\eea
Comparing \rf{A1} and \rf{A2}, we see that
\beq
 <\prod_i \z[U(C_i)]> \approx \prod_i <\z[U(C_i)]> 
\label{R2}
\eeq
The center elements of the holonomies therefore fluctuate independently.

   Any SU(2) class function $F[U(C_i)]$ can be expressed as a function of
of the sign and the modulus of Tr$[U(C_i)]$ as follows:
\beq
       F[U(C_i)] = F_1[\mbox{Tr}^2 \{U(C_i)\}] + 
                     \z[U(C_i)] F_2[\mbox{Tr}^2\{U(C_i)\}]
\eeq
VEVs of products of $F_1$ and $F_2$ do not factorize, as seen in
eqs.\ (\ref{F1}-\ref{F3}), while products
of the $\z[U(C_i)]$ do factorize, as seen above in \rf{R2}.
The conclusion is that there is magnetic disorder in the center 
elements $\z$ of the SU(2) group holonomies, but not in the coset elements, 
which depend only on $\mbox{Tr}^2[U(C)]$.

   All of this has some bearing on the question of what are the
relevant confining configurations in non-abelian gauge theory.
Confining configurations, whatever they may be in D=3 and D=4 dimensions,
must have the property of disordering the \emph{signs} of Wilson loops,
but not disordering the absolute values of those loops.  At least,
we have seen that this must be true at strong coupling.  A class of
configurations with these properties is the ``spaghetti vacuum'' of center
vortices, proposed twenty years ago, in various forms, by
't Hooft \cite{tHooft}, Mack \cite{Mack}, the Copenhagen group
\cite{Cop,jan,jo1,jo2}, and others \cite{Vinci,Corn1,Feyn}.  A center vortex,
linking loop $C$, has the property of sending $U(C) \ra z U(C)$;
such configurations can \emph{only} disorder the center elements
(i.e. the signs of Wilson loops, in SU(2)), leaving the rest
of the group distribution untouched.  This seems to be exactly
what is needed.

   On the other hand, it is not at all excluded that there could be 
some other degrees of freedom associated with loops that also fluctuate 
independently, and contribute to magnetic disorder.  In our discussion so 
far we have only considered the holonomies $U(C)$, but we could also 
imagine using, e.g., an adjoint Higgs field (either elementary or composite)
to construct other types of loop elements that might 
fluctuate independently.  In fact, this is the general idea behind
monopole confinement:  An adjoint Higgs field is used to single out
a U(1) subgroup of SU(2), and it is U(1) group elements, associated
with loops $C_i$, that are disordered via a monopole condensate.  From this 
point of view,  the $Z_2$ disorder could be just a subset of a more general 
U(1) disorder.  The simplest and most explicit proposal for monopole 
confinement (and U(1) disorder) in a non-abelian gauge theory is due to 
Polyakov, in his analysis of the D=3 Georgi-Glashow model \cite{Poly,book}.  
Since $GG_3$ interpolates smoothly between $QED_3$ and pure $YM_3$, we
would now like consider if there is any remnant of $Z_2$ disorder in the
3D Georgi-Glashow model.
 
\section{Double-Charged Loops in $\mathbf{GG_3}$}

The full lattice
action of the Georgi-Glashow model is a function of the gauge field variables
$U_\m(x)$ and the adjoint Higgs field variables $\phi(x)$ 
\bea
     S[U,\phi] &=& \oh \b_G \sum_{plaq} \mbox{Tr}[UUU^\dg U^\dg]
\non \\
       &+& \oh \b_H \sum_{x,\m} \mbox{Tr}[U_\m(x)\phi(x)
               U_\m^\dg (x) \phi^\dg(x+\hat{\m})]
\non \\
       &-& \sum_x \left\{ \oh \mbox{Tr}[\phi \phi^\dg]
                 + \b_R \left( \oh \mbox{Tr}[\phi \phi^\dg] - 1
                        \right)^2 \right\} 
\label{GG3}
\eea
where
\beq\label{ja0}
       \phi(x) = \sum_{a=1}^3 \phi^a(x) \s_a
\eeq

The naive continuum limit is obtained from the lattice 
action by the following scaling:
\beq\label{ja1}
U_\m(x)= e^{i a A_\m(x)},~~(A_\m\equiv \oh\s^a A_\m^a),~~~~~~
\phi= a^{1/2} \b^{1/2}_R \varphi,
\eeq
and the identification of continuum couplings $g$, $\l$ and $\m$ by
\beq\label{ja2}
\b_G = \frac{4}{g^2a},~~~\frac{\b_R}{\b_H}=\oq \l\,a,~~~
\frac{1-3\b_H-2\b_R}{\b_H}= - \oh \m \, a^2. 
\eeq
Thus the continuum action becomes\footnote{A proposal for the
effective action in dual variables is found in ref.\ \cite{Kov}.}
\beq\label{ja3}
S_{cont}= -\int d^3x\; \oh{\rm Tr}\, \Bigl[\frac{1}{g^2}\,F_{\m\n}^2+
\oh\Bigl(D_\m \varphi\Bigl)^2 -\oh\m \,\varphi^2 
+\oq\l\, \varphi^4\Bigr],
\eeq
where $D_\m\varphi = \pa_\m \varphi +i[A_\m,\varphi]$. 
It is obtained for 
\beq\label{ja4}
\b_G \to \infty,~~~\b_R \to 0,~~~\b_H \to \frac{1}{3},
\eeq
the approach monitored by the lattice spacing $a$. 
A more precise description is obtained by taking into account
that although the coupling constants are not renormalized in three 
dimensions, the masses $M_W$ and $M_\varphi$ {\em are} renormalized.
We refer to \cite{Teper} for formulas which include the one-loop 
mass renormalizations. The tree-formulas describe correctly the 
qualitative features which have our interest. 
In the ``broken''
region, where the approach of $\b_H$ to 1/3 is from above, it 
corresponds mass $M_W$ for the charged vector particle and 
a mass $M_\varphi$ for the neutral scalar, given by:
\beq\label{ja5}
M^2_W=g^2\,\varphi_c^2,~~~M_\varphi^2= 2\m \,\varphi^2_c,~~~~
\varphi_c^2 = \frac{\m}{\l}.  
\eeq

If we in the limit \rf{ja4} increase $\b_H$ away from 1/3, it corresponds 
formally to a ``continuum'' limit where
\beq\label{ja6}
M_W \sim \frac{1}{a},~~~~M_\varphi \sim \frac{1}{a},
\eeq
i.e.\ a limit where only the free photon field is present 
as a physical excitation. More generally, taking $\b_G$ and $\b_R$ fixed
and $\b_H \to \infty$ have the same effect, except that the resulting 
effective theory will be compact lattice $U(1)$, where we obtain  
the continuum free photon field only in the limit $\b_G \to \infty$.
For future reference let us note that the tree-value formulas \rf{ja2} and 
\rf{ja5} lead to the following expressions for dimensionless ratio 
\beq\label{ja7}
\frac{M_W}{g^2} = \frac{\b_H \b_G}{4} 
\left(\frac{3-(1-2\b_R)/\b_H}{2\b_R}\right)^{\oh}.
\eeq 
The actual region above $\b_H = 1/3$ where $M_W$ is not of the order 
of the inverse lattice spacing $a$ is very narrow, in agreement with 
numerical simulations \cite{Teper}.

Going to the unitary gauge, we write
\beq
      S_{ug}[U,\rho] \equiv S[U,\phi(x)=\rho(x)\s_3]
\eeq
The unitary gauge action $S_{ug}$ still has a residual $U(1)$
gauge symmetry.  We may factor the SU(2) link variable into a matrix $A_\m$
which transforms under the residual symmetry like a $U(1)$ gauge field, and
a piece $C_\m$ transforming like a double-charged matter field 
\cite{Kronfeld}
\bea
       U_\m(x) &=&  \left( \begin{array}{cc}
                      \cos\vph e^{i\th}  & \sin\vph e^{i\chi} \cr
                     -\sin\vph e^{-i\chi} & \cos\vph e^{-i\th}
                     \end{array} \right)
\non \\
               &=&   \left( \begin{array}{cc}
                      \cos\vph   & \sin\vph e^{i\gamma} \cr
                     -\sin\vph e^{-i\gamma} & \cos\vph 
                     \end{array} \right)
                      \left( \begin{array}{cc}
                          e^{i\th}  & 0 \cr
                            0       & e^{-i\th} 
                      \end{array} \right)
\non \\
               &=&   \left( \begin{array}{cc}
                       \sqrt{1-cc^*}   &  c \cr
                        -c^*           & \sqrt{1-cc^*} 
                     \end{array} \right)
                      \left( \begin{array}{cc}
                          e^{i\th}  & 0 \cr
                            0       & e^{-i\th} 
                     \end{array} \right)
                =    C_\m(x) A_\m(x)
\eea
with lattice measure 
\beq
       \int dU = {1\over 2\pi^2} \int_0^{\pi/2} d\vph ~ \cos\vph \sin\vph
            \int^\pi_{-\pi} d\th  \int^\pi_{-\pi} d\chi
\eeq
The effective abelian action $S_{eff}[A]$ for $GG_3$ is then defined by
\beq
 e^{S_{eff}[A]} \equiv \int \prod_x d\rho(x) \prod_\m d\vph_{\m}(x) 
       d\chi_{\m}(x) \cos \vph_{\m}(x)  \sin\vph_{\m}(x) 
      ~ e^{S_{ug}[U,\rho]}
\label{Seff}
\eeq
    
   The question we wish to raise is whether the Euclidean quantum theory of 
$S_{eff}[A]$, at large scales, is correctly represented 
by a monopole Coulomb gas, as proposed in \cite{Poly}, since this is 
the generic model of monopole confinement. 
In particular, consider the ``abelian'' Wilson loops 
\beq
      A_n(C)  =  \mbox{Tr}[(AAA...A)^n]
\eeq
corresponding to closed loops of heavy particles carrying $n$ units
of the abelian electric charge, with
\beq
     <A_n(C)> = {1\over Z} \int DA ~ A_n(C) e^{S_{eff}[A]}
\label{Aeff}
\eeq 


  The crucial point is that in $GG_3$, the string-tension $\s_n$ of
loops $<A_n(C)>$ should vanish asymptotically, if $n$ is an even integer.
The reason is simply that the Georgi-Glashow model contains massive
W-bosons which carry two units of electric charge; these 
correspond to the $C_\m(x)$ degrees of freedom.
The W-bosons are able to screen static sources carrying 
an even number of unit electric charges. If a flux tube were to
form between $n$=even charged sources, with some string tension
$\s_n=T$ then at a separation of roughly
\beq
         L = {n M_W \over T}  
\label{screen}
\eeq
charge-screening by W-bosons becomes energetically favorable and the 
flux-tube breaks, so $\s_n=0$ asymptotically.  Although $S_{eff}[A]$ contains 
only the abelian gauge field, it must somehow incorporate this non-confinement
of even charges, since we can always rewrite eq. \rf{Aeff} as
\beq
     <A_n(C)> = \int DA  D\vph D\chi D\rho ~  A_n(C) e^{S_{ug}[C_\m A_\m,\rho]}
\eeq
where the charged fields are included explicitly.

\subsection{Strong coupling expansion}

While charge-screening may be expected in $GG_3$ on very general grounds,
it is also possible to verify the effect explicitly in a strong-coupling
expansion.  Take, for simplicity, $\b_R=\infty$ so that the modulus
of the Higgs field is ``frozen'' to $\rho=1$, with $\b_H$ chosen large enough
such that $M_W \gg 1$ in lattice units, and $\b_G$ small enough to
allow a strong-coupling expansion.  Expanding the action to 2nd order
in $c,~c^*$, one easily finds for large, double-charged loops the perimeter
falloff expression
\beq
      <A_2(C)> = \exp[-\m \P(C)]  ~~~~~~(\mbox{GG}_3)
\eeq
with perimeter coefficient, extracted from the leading diagram,
\beq
        \m = - \log \left[ {\b_G^2 \over 8\b_H} { 1 - 2\b_H e^{-2\b_H}
                  - e^{-2\b_H} \over 1 - e^{-2\b_H} } \right]
\eeq
In strong-coupling compact QED, of course, the answer is different.
There are no charged bosons to screen the double-charged loop, and
its value, for the Wilson action, is
\beq
  <A_2(C)> = \exp[-2\s \A(C)] ~~~~~~ (QED_3)
\label{QED3}
\eeq
with $\s$ the string tension of the single-charged loop.
For multiply charged loops $<A_n(C)>$ in strongly-coupled
compact $QED_3$, the string tension is in general $n$ times the string
tension for single-charged loops.  Thus we have a qualitative difference,
at least in strong-coupling, between compact $QED_3$ and $S_{eff}[A]$ of
the 3D Georgi-Glashow model, because in the latter effective abelian theory,
there is no string tension for $<A_n(C)>$ when $n$ is even.

   For compact $QED_3$, it is possible to derive the result \rf{QED3}
also at weak couplings, which we do in the next section.  This is an 
interesting result in its own
right, since the existence of the $n=2$ string tension in compact
$QED_3$ has been questioned (ref. \cite{book}, p. 80), while, 
in numerical simulations, a finite value equal to twice the $n=1$ value
has been measured \cite{Zach}.

\subsection{Double-Charged Loops in a Monopole Gas}

It is well known that compact $U(1)$ on a lattice 
can be written in a monopole gas represention.
If we use the Villain version of compact $U(1)$ one obtains
\cite{bmk}
\beq\label{j0}
Z_{mon} = \sum_{m(r)=-\infty}^{\infty} 
\exp\Bigl[ -\frac{2\pi^2}{g^2a} \sum_{r,r'}  
m(r') G(r-r') m(r) \Bigr],
\eeq
where $m(r)$ is an integer valued monopole field at the (dual) lattice size 
$r$, $G(r-r')$ is the lattice Coulomb  propagator in three
dimensions, i.e. $\Delta_\m^2G(r-r')=-\delta_{rr'}$, 
and $g^2a$, $a$ being the lattice spacing,  is the 
temperature $1/\beta$ in the usual thermodynamic interpretation.

One can represent the propagator $G(r-r')$ by a Gaussian 
functional integral:
\bea\label{j0a}
Z_{mon} &=& \int \prod_r d\chi(r) \; \exp\Bigl[-\frac{g^2a}{4\pi^2} 
\sum_{r} \oh (\Delta_\m \chi(r))^2\Bigr] 
\non \\
        & & \times \sum_{m(r)=-\infty}^\infty
        \exp\Bigl[-{2\pi^2 \over g^2 a}G(0)\sum_r m^2(r) 
        + i\sum_r m(r) \chi(r) \Bigr].
\eea
In the weak coupling limit (low temperature limit)
where $g^2a \to 0$ we need only 
to maintain the first terms $|m(r)| \leq 1$ in the sum and we get
\beq\label{j1a}
Z_{mon} \approx \int \prod  d\chi(r) \; \exp \Bigl[-\frac{g^2a}{4\pi^2} 
\sum_r\Bigl( \oh (\Delta_\m \chi)^2 +M_0^2(1-\cos \chi(r)) \Bigr)\Bigr],
\eeq
where $M^2_0$ comes from the propagator $G(r-r')$ for coinciding 
arguments, ($G(0) \approx 0.253$ \cite{watson})
\beq\label{j1b}
M^2_0 = \frac{8\pi^2}{g^2a}\, \exp\Bigl[-\frac{2\pi^2}{g^2a} \,G(0)\Bigr].
\eeq
and if we interpret the monopoles as instantons,  $2\pi^2G(0)/g^2a$ can 
be viewed as the action of the instanton. 

Let us for notational simplicity carry out the following 
discussion in a continuum notation. All manipulations done
in the following have a precise lattice analogy (see \cite{bmk}).
The translation is 
\beq\label{jzz}
\Delta_\m\to a\,\pa_\m,~~~a^3 \sum_r \to \int d^3 r ,
\eeq
and we end up with 
\beq\label{j1}
Z_{mon} \approx \int \cD \chi(r) \; 
\exp \Big[ -\frac{g^2}{4\pi^2} 
\int d^3r \;\Bigl( \oh (\pa_\m \chi)^2 +M^2(1-\cos \chi(r)) \Bigr)\Bigr].
\eeq
where 
\beq\label{j1d}
M^2=a^{-2}M_0^2.
\eeq

In the  Coulomb gas approximation we are effectively integrating
over the electromagnetic fields carried by the monopoles.
They can be found from the monopole density $m(r)$ by integrating
\beq\label{j4}
\pa_\m H_\m(r) = 2\pi \, m (r),~~~{\rm i.e.}~~~ 
H_\m(r) = \oh\int d^3r' \;\frac{(r-r')_\m}{|r-r'|^3}\; m(r'),
\eeq
In particular we find, if $C$ denotes a closed curve and $S(C)$
a surface with $C$ as boundary, that 
\beq\label{j6}
\oint_C dr_\m A_\m(r) = \int_{S(C)} dS_\m(r)\; H_\m(r) =
\int d^3r \;\eta_{S(C)}(r)\; m(r),
\eeq
where
\beq\label{j7}
\eta_{S(C)}(r) = 
-\oh \frac{\pa}{\pa r_\m} \int_{S(C)} dS_\m(r') \;\frac{1}{|r-r'|}.
\eeq
It is seen that $\eta(r)$ has the interpretation as a dipole sheet 
on the surface $S(C)$. Thus, if $C$ is a planar curve in the 
$(x,y)$ plane and $S(C)$ the planar surface with $C$ as boundary curve,
we have 
\beq\label{j7a}
-\pa^2 \eta_{S(C)} = 2\pi \delta'(z) \theta_{S}(x,y),
\eeq
where $\theta_{S(C)}(x,y)$ is one for a point inside the boundary $C$ and 
zero for a point outside the boundary $C$. Close to the surface $S$ we have 
\beq\label{j7b}
\eta(r) = \pi {\rm sign}\, z \; \theta_{S}(x,y),
\eeq
i.e.\ it jumps by $2\pi$ when passing the dipole sheet.

We can now calculate the expectation value of a planar Wilson loop
which carries n units of charge. From \rf{j6} we obtain
\beq\label{j2}
\lan A_n(C)\ran \equiv \lan e^{in \oint dr_\m\; A_\m(r)} \ran = 
\lan e^{ i n\int d^3r \;\eta_{S(C)}(r)\;m(r)}\ran,
\eeq
where the expectation value is calculated with respect to 
the partition function \rf{j0}.   
Performing the same transformations which lead from \rf{j0}
to \rf{j1}, the last term in \rf{j2} leads to a translation
of $\chi(r)$ such that one obtains 
\beq\label{j8}
\lan A_n(C)\ran = 
\frac{1}{Z_{mon}} 
\int \cD  \chi(r) \;  
\exp \Bigl[-\frac{g^2}{4\pi} 
\int d^3r \;\Bigl( \oh (\pa_\m (\chi-n 
\eta_{S(C)})^2 +M^2(1-\cos \chi(r)) \Bigr)\Bigr],
\eeq

Since we consider the weak coupling regime the dominant 
contribution to the expectation value \rf{j8} comes from the 
classical solution to the effective action in \rf{j8}.
In case we choose the planar loop to lie in the $(x,y)$ plane
we obtain from \rf{j7a}:
\beq\label{j9}
\pa^2 \chi = 2\pi n \delta'(z)\theta_S(x,y) + M^2 \sin \chi.
\eeq
A solution to the homogeneous equation is (far from the boundary,
suppressing a trivial $(x,y)$ dependence):
\beq\label{j14}
\chi^{(0)}(z)=4\arctan e^{-Mz}. 
\eeq 
Note that for $z < 0$ it can be written as $-4 \arctan e^{Mz}+2\pi$.
Thus, for $n=1$, eq.\ \rf{j9} has, 
far away from the boundary $C$,  the solution 
\beq\label{j10}
\chi_{cla}^{(1)}= {\rm sign}\, z\cdot 4 \arctan (e^{-M\,|z|}) \, \theta_S(x,y).
\eeq
The important property of the solution \rf{j10} is that 
$\chi_{cla}^{(1)}(z) \to 0$ for $|z| \to \infty$. This implies 
that it can be joined to the trivial solution $\chi=0$ for $|r| \to 
\infty$ in $R^3$.

Since the solution $\chi_{cla}^{(1)}$ is given in terms of elementary 
functions one can calculate the corresponding action in \rf{j8}:
\beq\label{j11}
\int d^3x \Bigl( \oh (\pa_\m (\chi_{cla}^{(1)}- 
\eta_{S(C)})^2 +M^2(1-\cos \chi^{(1)}_{cla}(x)) \Bigr) = 8M \, {\rm Area}(S)+ 
{\rm perimeter~contributions} .
\eeq
In this way one obtains the famous area law of Wilson loops \cite{Poly}
in the three dimensional Coulomb gas of monopoles since 
it follows from \rf{j8} that 
\beq\label{j12}
\lan A_1 (C) \ran \approx \exp\Bigl[ - \frac{g^2}{4\pi^2} \,8M\;  
     {\rm area}(S)\Bigr].
\eeq

Let us now consider the situation for double charged Wilson loops.
In order to find the minimum of the action we should solve 
\rf{j9} for $n=2$. One could be misled to suggest the simple solution
(far from the boundary of $C$)
\beq\label{j13}
\chi_{cla}^{(2)}(z) = 2\pi {\rm sign}\, z,~~~{\rm i.e.}~~~\chi(z)=2\eta(z).
\eeq
This clearly gives energy zero in the interior of the Wilson 
loop and seemingly no area law (as for the full Georgi-Glashow 
model). However, since this solution does not go to zero 
far from the Wilson loop, contrary to the solution \rf{j10}
for $n=1$, we have to interpolate between  
the limiting values $\chi_{cla}^{(2)}(z)=2\pi$ and 
$\chi_{cla}^{(2)}(z)=-2\pi$ in some 
region in space far away from the Wilson loop.  
This will cost an energy which is easily seen to be 
proportional to the area of the ``domain wall'' where the interpolation
takes place. Thus the optimal situation is also here one where 
$\chi_{cla}^{(2)}(z) \to 0$ for $|z| \to \infty$. With this requirement 
it follows that we have to solve eq.\ \rf{j9} with 
the boundary conditions that $\chi_{cla}^{(2)}(z) \to 0$ for 
$|z| \to \infty$ and $\chi_{cla}^{(2)}(z)-2\eta(z)$ is differentiable 
for $z=0$ when passing the sheet.  

There is no such solution. But we can find approximate solutions
with energies above, but arbitrary close to twice the energy
corresponding $\chi_{cla}^{(1)}$. From \rf{j9}, \rf{j10} and \rf{j14} 
it follows that for $z_0 \gg 1/M$ 
\beq\label{j16}
\chi^{(2)}(z)= \theta(z) \, \chi^{(0)}(z-z_0)+ 
\theta(-z)(\chi^{(0)}(z+z_0)-2\pi)
\eeq
is a solution to \rf{j9} with $n=2$ except for exponentially 
small corrections, and its energy is twice that of $\chi^{(1)}_{cla}$
except for exponentially small corrections. Further we see that this 
approximate solution behaves like \rf{j13} for $|z| << z_0$.

It is interesting that one can find a different kind of solution
with the same features, namely that the energy of the classical solution 
can be arbitrary close to twice that of $\chi_{cla}^{(1)}$, 
but never reach it. We have a free choice for the surface $S(C)$,
except for the requirement that $C$ is the boundary of $S$. In particular,
we could for the doubled charged Wilson loop choose {\it two} 
sheets separated a distance $2d$  and located in the $z= \pm d$-planes,
except close to the boundary $C$. For each sheet we now 
have a discontinuity corresponding to $\eta(z)$. If $d >> 1/M$ 
it is clear that the solution, except for exponentially small corrrections,
has to be
\beq\label{j17}
\chi^{(2)}(z) = \theta(-z) \chi^{(1)}_{cla}(z+d)+
\theta(z)\chi^{(1)}_{cla}(z-d)W.
\eeq
The energy becomes minimal (and equal two times that corresponding 
to $\chi^{(1)}_{cla}(z)$) in the limit $d \to \infty$.

We conclude that the string tension for a double-charged Wilson 
loop will be twice the string tension of a single charged 
Wilson loop if we restrict ourselves to the Coulomb gas approximaton of 
functional integral. Clearly the arguments can be extended to 
$n$-charged Wilson loops.
 
\subsection{Weak coupling limit of lattice $\mathbf{GG_3}$}

The monopole gas calculation above was a {\em weak coupling expansion}
in the sense that $g^2a$ had to be considered small in 
order to make the truncation \rf{j1a}. In particular this implied 
that the monopole ``action'' $2\pi^2G(0)/g^2a$ is large and the 
density of monopoles,
\beq\label{j20}
\rho \sim \exp\Bigl(-\frac{2\pi^2G(0)}{g^2a}\Bigr), 
\eeq
exponentially small. In the  naive continuum limit, as defined by
\rf{ja2}-\rf{ja5}, we can make contact to the similar instanton 
calculations in the continuum $GG_3$ model. In that case 
the instanton (monopole) action is given by
\beq\label{j21}
S_{mon} = \frac{M_W}{g^2}\, \epsilon(\l/g^2),
\eeq
where $\epsilon(x)$ is a slowly  varying function of $x$ ($\epsilon(0)=4\pi$).
(It is seen that one obtains the $QED_3$ formulas in the limit 
where $M_W \sim 1/a$, as expected). 
A dilute instanton calculation is valid if the density of instantons, 
\beq\label{j22}
\rho \sim \exp(-S_{mon}),
\eeq
is exponentially small relatively to the  extension of the instantons 
(which is $\approx 1/M_W$). Thus the calculation in the last 
subsection is valid in $GG_3$ provided $M_W/g^2 \gg 1$. One 
obtains a string tension for an $n$-charged Wilson loop
\beq\label{j23}
T_n \sim n\; e^{-S_{mon}}\, \times\,[\mbox{subleading corrections}].
\eeq 

To the extent one can trust the tree-value formulas, the  
lattice inequality corresponding to $M_W/g^2 \gg 1$ can be obtained from 
\rf{ja7}:
\beq\label{j24}
\frac{M_W}{g^2} \sim \frac{\b_H \b_G}{4} 
\left(\frac{3-(1-2\b_R)/\b_H}{2\b_R}\right)^{\oh} \gg 1.
\eeq  
This formula is most 
reliable for $\b_G$ large (and $\b_R$ is small) and $\b_H$ 
close to $(1-2\b_R)/3$. 

We have seen above (see eq.\ \rf{screen}) that we expect a perimeter 
law in $GG_3$ for $n$-charged Wilson loops, $n$ even, provided the 
linear extension $L$ of the loop satisfies 
\beq\label{j25}
L > \frac{n M_W}{T_n} \sim e^{S_{mon}}\times[\mbox{subleading corrections}]
\eeq
This length is much larger than the length scale
\beq\label{j26}
\xi_\s = \s^{-1/2} \sim \exp\Bigl(\oh S_{mon}\Bigr),
\eeq
set by the string tension, and it is also much greater than the 
average distance 
\beq\label{j27}
R = \rho^{-1/3} \sim \exp\Bigl(\frac{1}{3} S_{mon}\Bigr),
\eeq
between the monopoles. Typically we will have to 
go to distances larger than $\xi_\s$ if we want to 
measure the string tension. However, for $M_W/g^2 \gg 1$ 
we have to move out exponentially many units of length $\xi_\s$
before an $n$-charged string, $n$ even, breaks:
\beq\label{j28}
\frac{L}{\xi_\s} \sim \exp \Bigl( \oh S_{mon} \Bigr) \times 
[\mbox{subleading corrections}].
\eeq

In the lattice $GG_3$ model \rf{GG3} the parameter $\b_H$ allows
us to interpolate continuously between compact $QED_3$ (large $\b_H$)
and pure Yangs-Mills theory ($\b_H \to 0$). From the tree-formula 
\rf{j24} we see (for large $\b_G$) how large $\b_H$ corresponds to 
a value $M_W/g^2 \gg 1$, while $M_W/g^2$ 
drops to zero (in the tree-approximation) for $\b_H =(1-2\b_R)/3$.
Below this value of $\b_H$ we have the ``unbroken'' coupling region of the
Yang-Mills-Higgs system, where we expect the center $Z_2$ to play the 
dominant role in confinement and the monopole gas description is not valid 
at all.

\section{$\mathbf{Z_2}$ Disorder in $\mathbf{GG_3}$}

   We now return to the question of $Z_2$ disorder in $GG_3$.
Considering only the abelian magnetic flux probed by loops $A_n(C)$,
we can ask if disorder is distributed evenly in the $U(1)$ group,
or if it is only present in some subset of the degrees of freedom.  Our 
procedure is the same as in section 3.  Defining again the holonomy
distributions on the compact $U(1)$ group
\bea
        P_C(e^{i\w}) &=& <\d[\w,\th(C)]> = 
                <\sum_{n=-\infty}^{\infty} e^{in[\w - \th(C)]}>
\non \\
      \tP_C(e^{i\w}) &=&  P_C(e^{i\w}) - 1
\eea
we see that for compact $QED_3$ the approach to a pure random distribution 
has an area-law falloff 
\beq
      \tP_C(e^{i\w}) \sim \cos(\w) e^{-\s \A(C)}  ~~~~~ (QED_3)
\eeq
while in $GG_3$, the approach goes, asymptotically, as a perimeter-law 
falloff 
\beq
      \tP_C(e^{i\w}) \sim \cos(2\w) e^{-\m \P(C)}  ~~~~~~ (GG_3)
\eeq
due to the different behavior of the $n=$even charged loops.  However,
once again, there is a hidden area-law approach to randomness also in
$GG_3$, since if we define 
\beq
         \z[A(C)] = \mbox{sign}\cos[\th(C)]
\eeq
and the probabilities
\beq
       P_C(z) = <\oh (1 + z \times \mbox{sign}[\cos\th(C)])>
\eeq
with $z=\pm 1$ and $\tP_C(z)=P_C(z)-\oh$, then in $GG_3$
\beq
      \tP_C(z)  \sim  z e^{-\s \A(C)}
\eeq

   Turning to lattice strong coupling we again find, in complete
analogy to the pure-gauge theory in section 3, that at leading
order for compact $QED_3$
\beq
     <\prod_i F[A(C_i)]> \approx  \prod_i <F[A(C_i)]>  ~~~~~ (QED_3)
\eeq
while in $GG_3$
\bea
     <\prod_i F[A(C_i)]> &\sim&  e^{-\m \P(C)}
\non \\
     \prod_i <F[A(C_i)]> &\sim&  e^{-\m \sum_i \P(C_i)}
\eea
It follows that, in contrast to $QED_3$, the abelian loop elements do not 
fluctuate independently in $GG_3$, even at very strong lattice coupling.
On the other hand, the $Z_2$ elements
\bea
      <\prod_i \z[A(C_i)] > &=& \prod_i  e^{-\sum_i \A(C_i)}
\non \\
            &=& \prod_i <\z[A(C_i)]>
\eea
\emph{do} fluctuate independently in $GG_3$, at strong coupling.

   The conclusion is that even in $GG_3$, long-range disorder seems to
be associated with a $Z_2$, rather than a $U(1)$, subgroup; there is again
disorder in the sign, but not in the modulus, of loop elements $\cos\th(C)$.
The inclusion of an adjoint Higgs field does not seem to change the fact
that disorder, at large scales, is essentially a property of the gauge group
center. In the last section we gave a qualitative description of 
the length scales in $GG_3$ beyond which the Coulomb gas picture 
breaks down and where (as we have now argued) the magnetic disorder 
is center disorder.

\subsection{Extension to SU(N)}

   All of the arguments above are readily extended to theories with
an SU(N) gauge group; we will only indicate briefly how this goes.
The probability distribution $P_C(g)$ in eq.\ \rf{prob} generalizes 
in the obvious way, with the sum over $j$ replaced by a sum over $SU(N)$ 
representations.  Writing
\beq
           \chi_F(g) = A e^{i\phi}
\eeq
where $F$ denotes the fundamental representation, $A \ge 0$ is real,
and $\phi \in [0,2\pi)$, let
\bea
          n(g) &=& \mbox{int}\left[{N \phi \over 2\pi} \right]  
\non \\
          \z(g) &=& \exp[2\pi i n(g)/N]
\eea
where int($x$) denotes the integer part of the real number $x$.  This
definition assigns a center element to every group element, 
with the property that $\z(zg)=z\z(g)$ for $z \in Z_N$.  Then
\beq
  P_C(z) = <\Phi[z,U(C)]> ~~~ \mbox{where} ~~~     
         \Phi[z,U(C)] = \left\{ \begin{array}{cl}
                       1 & \mbox{if~} z = \z[U(C)] \cr
                       0 & \mbox{otherwise} 
                             \end{array} \right.
\eeq
gives the probability that $\z[U(C)]=z$.

   Arguments entirely analogous to those in sections 2 and 3 show
that the holonomy probability $P_C(g)$ approaches the random distribution
with only a perimeter-law falloff, while $P_C(z)$ approaches the
random distribution with an area-law falloff.  At strong-couplings,
the center elements $\z[U(C_i)]$ fluctuate independently in sub-areas of a 
large loop, while class functions $F[U(C_i)]$ in general do not.  From this 
we conclude that there is magnetic disorder among the center elements, but 
not in the coset. Once again, it should be noted that the 
correlation among $SU(N)/Z_N$ coset elements relies on non-planar 
contributions, which are dominant for large loops.  If we would take the 
large-N limit before the large-loop limit, then there is Casimir scaling as 
in D=2 dimensions, and disorder throughout the group manifold.

   Introducing an adjoint Higgs field in $D=3$ dimensions, and fixing to
unitary gauge, we can define the loop observables (in continuum notation),
invariant under the remnant $U(1)^{N-1}$ subgroup
\beq
 <rk|A(C)|rk> = \left[ \exp[i\oint_C dx^\m ~ A_\m^i(x) H^r_i] \right]_{kk}
\eeq
where $H^r_i$ denotes the $i$-th generator of the Cartan subalgebra in
representation $r$ of the SU(N) group, and $(kk)$ is just an
element of the dim($r$)$\times$dim($r$) diagonal matrix $A(C)$ in this
represention.  The $Z_N$ center elements $\z[A(C)]$ are extracted, as above, 
from the phase of $\chi_F[A(C)]$.  

   In the monopole Coulomb gas picture, disregarding the effects of
the charged bosons, the string tension of $<rk|A(C)|rk>$ depends on
both representation $r$ and choice of diagonal matrix element $(kk)$.  
Allowing, however, for the effects of the charged W-bosons, these string 
tensions can depend only
on the N-ality of representation $r$, and are independent of $(kk)$.
Asymptotically there is disorder in the $\z[A(C)]$ elements, but not in the
full $U(1)^{N-1}$ group manifold.  The distribution of abelian magnetic
flux, in the $Z_N$ disorder regime, is not that of a monopole Coulomb
gas.

\section{Discussion}
   
   We have stressed in this article the fact that, while
massive virtual particles are often irrelevant to vacuum fluctuations
in the far-infrared, this is not the case for massive charged particles 
in a confining theory.  The screening of external charged sources by
quanta of the matter field is, of course, a rather trivial point, and
allows us to conclude that certain loop operators have a perimeter 
falloff.  What may be slightly less obvious is the fact that such
perimeter falloffs have implications for the probability distribution of 
large-scale vacuum fluctuations also in the \emph{absence} of external 
charges.  This point is best appreciated in, e.g., the 3D Georgi-Glashow
model, by imagining an integration, in unitary gauge, over the W-bosons 
and Higgs field, to leave an effective action $S_{eff}[A]$ involving
only the photon field.  There are no longer any explicit, electrically 
charged fields left in the action to screen multiply-charged abelian loops.  
Instead, the effect of the virtual W-particles has gone into altering
the Boltzman distribution for vacuum fluctuations of the abelian field,
such that those abelian configurations which would lead to an area law
for even-charged loops in the $Z_2$ regime have been 
suppressed.  Thus the effective action $S_{eff}[A]$ is not only 
quantitatively but also \emph{qualitatively} different, at large scales, 
from the $QED_3$ action with a lattice cutoff, and a monopole Coulomb
gas picture is not adequate to describe the confining vacuum in the 
$Z_2$ regime. 

\begin{figure}[h]
\centerline{\scalebox{.5}{\rotatebox{270}{\includegraphics{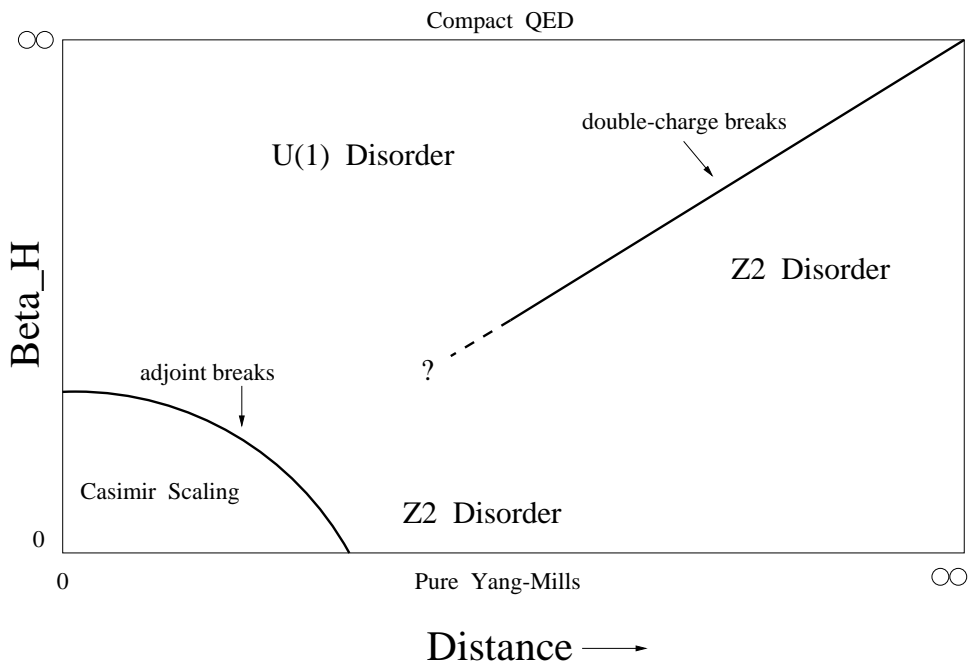}}}}
\caption{Confining disorder, extracted from U(1) and SU(2) holonomies, in the 
3D Georgi-Glashow model.}
\label{disorder}
\end{figure}

   The picture we are led to, for the onset of $Z_2$ disorder in the 3D 
Georgi-Glashow model, is indicated schematically in Fig.\ \ref{disorder}.
For fixed $\b_G$ and sufficiently large $\b_R$, no phase transition
is encountered as $\b_H$ varies from $\b_H=0$ ($YM_3$) to $\b_H=\infty$
($QED_3$) \cite{Teper}.  The curved solid line, however, represents the 
breaking of the adjoint string, and the loss of ``Casimir scaling,''  while 
the solid line tailing off in a dashed line represents the breaking of the 
flux tube between double-charged abelian sources.  The dashed line 
indicates the complete breakdown of the Coulomb gas picture in unitary gauge, 
as $\b_H \ra 0$.  All abelian Wilson loops vanish in this gauge in the 
$\b_H=0$ limit, although it may still be possible to define the
$n=2$ abelian charge screening distance by 
extrapolation from non-zero $\b_H$.  It would be interesting to know
where the dashed line terminates.   

   The absence of an adjoint string tension at any
length scale, for sufficiently large $\b_H$ at fixed ($\b_G,\b_R$), has been 
seen in numerical simulations of $GG_3$ \cite{Cas1,Jung}, and is easily 
understood.  A $j=1$ representation quark consists of two components
($m=\pm 1$) which are double-charged under the U(1) subgroup, and one 
component ($m=0$) which is 
neutral. When the confining field is essentially abelian, the neutral 
component is dominant, and the adjoint loop has no area-law falloff at any 
length scale.  In fact, this gives us an interesting criterion for U(1) 
disorder in an SU(2) gauge theory, regardless of whether the adjoint scalar 
is elementary or composite.  It is required that in the U(1) regime  
\begin{enumerate}
\item Even-charged loops have area-law falloff. Otherwise, as we have
seen, the loops are probing $Z_2$ disorder.
\item Adjoint loops have perimeter-law falloff.  If not, then
abelian neutral components are also subject to a confining force, and there
is disorder over the entire group manifold; not only in a U(1) or $Z_2$
subgroup.
\end{enumerate}
In D=4 dimensions, the maximal abelian gauge has been studied extensively
in pure Yang-Mills theory.  This gauge defines a composite adjoint Higgs 
field, U(1) holonomies, and monopole currents. The hope is that confining 
disorder is U(1) disorder which can be attributed, as in $GG_3$, to monopoles.
Numerically, however, although double-charged loops defined in this 
formulation have an area-law falloff (cf.\ ref.\ \cite{Poulis}) at the length 
scales probed by Monte Carlo simulations, this is also true of the adjoint 
loops at the same distance scales; the second criterion above is not
satisfied.

   Returning to the 3D Georgi-Glashow model at large $\b_H$, it is interesting
to consider how the confining abelian fields are arranged at large scales,
where there is $Z_2$ disorder.  It is useful to think in terms of the
effective abelian theory in \rf{Seff}, obtained from $GG_3$ in unitary gauge
by integrating out the W and Higg fields.  $S_{eff}[A]$ and $S[U,\phi]$ are 
of course equivalent, in unitary gauge, so far as the vacuum distribution of 
the A-field is concerned.
$S_{eff}[A]$, like the $GG_3$ model from which it is obtained, will have
instanton solutions corresponding to monopoles.  However, since the monopole
Coulomb gas picture breaks down at the onset of $Z_2$ disorder,
it must be that the interactions among monopoles are not really
Coulombic at long distances, and neither is the field distribution of
the corresponding magnetic flux.  This raises the interesting (although at this
stage speculative) question of how the abelian flux from monopoles is 
actually organized, on distance scales characteristic of the $Z_2$ regime.

\begin{figure}[h]
\centerline{\scalebox{.5}{\includegraphics{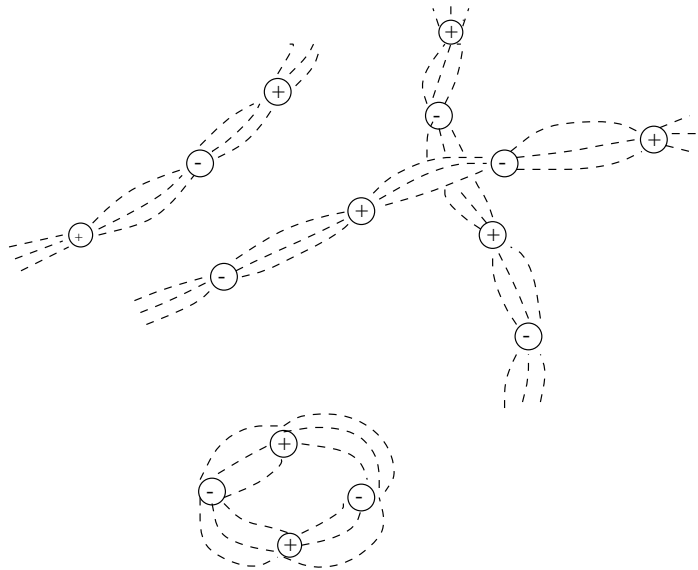}}}
\caption{An example of monopole-antimonopole magnetic flux
organized into $Z_2$ vortices.}
\label{vort6}
\end{figure}

  As it is only the sign of $\cos \th(C)$ which is disordered in the 
$Z_2$ regime, while the effective action $S_{eff}[A]$ only involves an abelian
gauge field, there is a strong implication that the magnetic flux due to 
monopoles is collimated, at sufficiently large distance scales, in units of 
$\Phi_B = \pm \pi$.  Collimated flux of these units, with a stochastic 
distribution of such ``fluxons'' across the minimal area of a large loop, 
affects only the sign of odd charged loops, leading to an identical 
string tension for all odd-charged loops, and yielding zero string tension 
for all even-charged loops. This is the proper result in the $Z_2$ disorder 
regime. If magnetic flux is, in fact, collimated in this way, then a 
$Z_2$ vortex picture in this regime is quite natural.  

   A particular example of $\pm 2\pi$ monopole flux organized into $Z_2$ 
vortex configurations of $\pm \pi$ flux is shown in Fig.\ 
\ref{vort6}.  This is by no means the only possibility.  In fact, 
at large $\b_H$, the scale $L$ at which monopole flux should be collimated 
in units of $\pm \pi$ is actually much greater than the average monopole 
separation $R$, as seen by comparing eqs.\ \rf{j25} and \rf{j27}. The 
illustration in Fig. \ref{vort6} might be relevant at lower $\b_H$, 
approaching the pure Yang-Mills limit, when $L/R$ 
is O(1).\protect\footnote{For a discussion of such 
configurations, in the context of the maximal abelian gauge in $YM_4$, see 
ref.\ \cite{Zako}.}  
As $\b_H \ra \infty$, the width of the vortex regions would
diverge to infinity, and the monopole Coulomb gas picture is valid
at all distances.  As $\b_H$ is reduced, the vortex width decreases. 
A very attractive possibility is that abelian vortices in $GG_3$ smoothly 
transform into center vortices of the pure Yang-Mills theory as
$\b_H \ra 0$, making contact with the ideas of refs.\ 
\cite{tHooft,Mack,Cop,jan,jo1,jo2,Vinci,Corn1,Feyn}, and the numerics of 
refs.\ \cite{Obs,TK,LR}.

\vspace{33pt}

\ni {\Large \bf Acknowledgements}

\bigskip

  We have benefited from discussions with Alex Kovner and Poul Olesen.

J.A. acknowledges the support of the Professor Visitante Iberdrola 
Program and the hospitality at the University of Barcelona, where part
of this work was done.  J.G.'s research was supported in part by
Carlsbergfondet, and in part by the U.S.\ Department of Energy under
Grant No.\ DE-FG03-92ER40711.


\begin{thebibliography}{xx}
\bibitem{tHooft}G. 't Hooft, Nucl. Phys. B138 (1978) 1. 
\bibitem{Mack} G. Mack, in {\sl Recent Developments in Gauge Theories},
edited by G. 't Hooft et al. (Plenum, New York, 1980).
\bibitem{Cop} H. B. Nielsen and P. Olesen, Nucl. Phys. B160 (1979) 380.
\bibitem{jan} J. Ambj{\o}rn and P. Olesen, Nucl. Phys. B170 (1980) 60; 265. 
\bibitem{jo1} J. Ambj\o rn and P. Olesen, 
 Nucl. Phys. B170 (1980) 265.
\bibitem{jo2}J. Ambj\o rn, B. Felsager and P. Olesen,
 Nucl. Phys. B175 (1980) 349. 
\bibitem{Vinci} P. Vinciarelli, Phys. Lett. 78B (1978) 485.
\bibitem{Corn1} J. M. Cornwall, Nucl. Phys. B157 (1979) 392.
\bibitem{Feyn} R. P. Feynman, Nucl. Phys. B188 (1981) 479.
\bibitem{Obs}  L. Del Debbio, M. Faber, J. Giedt, J. Greensite, and 
{\v S}. Olejn\'{\i}k, hep-lat/9801027.
\bibitem{TK} T. Kov\'{a}cs and E. Tomboulis, hep-lat/9709042;
hep-lat/9711009. 
\bibitem{LR} K. Langfeld, H. Reinhardt, and O. Tennert, hep-lat/9710068.
\bibitem{Cas} J. Ambj{\o}rn, P. Olesen, and C. Peterson, Nucl. Phys.
B240 [FS12] (1984) 198; 533; \\
C. Michael, Nucl. Phys. Proc. Suppl. 26 (1992) 417; Nucl. Phys. B259
(1985) 58; \\
N. A. Cambell, I. H. Jorysz, and C. Michael, Phys. Lett. B167 (1986) 91; \\
M. Faber and H. Markum, Nucl. Phys. Proc. Suppl. 4 (1988) 204; \\
M. M\"{u}ller, W. Beirl, M. Faber, and H. Markum, Nucl. Phys. Proc.
Suppl. 26 (1992) 423; \\
G. Poulis and H. Trottier, Phys. Lett. B400 (1997) 358, hep-lat/9504015.
\bibitem{Cas1} L. Del Debbio, M. Faber, J. Greensite, and 
{\v S}. Olejn\'{\i}k, Phys. Rev. D53 (1996) 5891.
\bibitem{Poly} A. Polyakov, Nucl. Phys. B120 (1977) 429.
\bibitem{book} A. Polyakov, {\sl Gauge Fields and Strings} (Harwood Academic
Publishers, 1987). 
\bibitem{Kov} A. Kovner and B. Rosenstein, Int. J. Mod. Phys. A8 (1993) 5575.
\bibitem{Teper} A. Hart, O. Philipsen, J. Stack, and M. Teper,
Phys. Lett. B396 (1997) 217, \\ hep-lat/9612021.
\bibitem{Kronfeld} A. Kronfeld. G. Schierholz, and U.-J. Wiese,
Nucl. Phys. B293 (1987) 461.
\bibitem{Zach} M. Zach, M. Faber, and P. Skala, hep-lat/9709017.
\bibitem{bmk}T. Banks, R. Myerson and J. Kogut, Nucl. Phys. B129 (1977) 493.
\bibitem{watson}G.N. Watson, Quart. J. Math. 10 (1939) 266.
\bibitem{Jung} C. Jung, hep-lat/9712025. 
\bibitem{Poulis} G. Poulis, Phys. Rev. D54 (1996) 6974, hep-lat/9601013.
\bibitem{Zako} L. Del Debbio, M. Faber, J. Greensite, and 
{\v S}. Olejn\'{\i}k, Proceedings of the Zakopane meeting
{\sl New Developments in Quantum Field Theory}, hep-lat/9708023.

\end{thebibliography}
\end{document}